\documentclass[prd,aps, preprintnumbers, showpacs, nofootinbib,superscriptaddress,notitlepage,twocolumn]{revtex4-2}
\usepackage[normalem]{ulem}
\usepackage{epsfig}
\usepackage{amsfonts}
\usepackage{amsmath}
\usepackage{slashed}
\usepackage{graphicx}
\usepackage{mathrsfs}
\usepackage{color}
\usepackage{mathtools}
\usepackage{simpler-wick}
\allowdisplaybreaks[4]

\begin{document}

%\preprint{APS/123-QED}

\title{Creating true muonium via charmonium radiative decay}% Force line breaks with \\

	\author{Jian-Ping Dai}
    \email{daijianping@ynu.edu.cn}
\affiliation{Department of Physics, Yunnan University, Kunming 650091, China}

	\author{Hai-Bo Li}
	\email{lihb@ihep.ac.cn}
\affiliation{Institute of High Energy Physics, Chinese Academy of Sciences, Beijing 100049, People’s Republic of China}
\affiliation{University of Chinese Academy of Sciences, Beijing 100049, People’s Republic of China}
  
	\author{Shuai Zhao}
	\email{zhaos@tju.edu.cn}
\affiliation{Department of Physics, Tianjin University, Tianjin 300350, China}

	\author{Zong-Ying Zheng}
	\email{Corresponding author. 3019227002@tju.edu.cn}
\affiliation{Department of Physics, Tianjin University, Tianjin 300350, China}

\date{\today}% It is always \today, today,

\begin{abstract}
True muonium, the bound state of a muon and an antimuon ($\mu^+\mu^-$), has long been theoretically predicted but remains experimentally elusive. We investigate the production of true para-muonium in the radiative decay of $J/\psi$ meson, and analyze the prospects for detecting true muonium in current and future high-energy $e^+e^-$ experiments, particularly focusing on the BESIII experiment and the proposed Super Tau-Charm Facility. Although the events are rare at the super tau-charm facility, the detection of true para-muonium via $J/\psi$ radiative decays could become feasible at its future updates.

%Our investigation assesses the discovery potential by considering both the theoretical cross-sections for true muonium production and the experimental sensitivities of these facilities.

%\begin{description}
%\end{description}
\end{abstract}

\maketitle

%\tableofcontents

\section{ Introduction }
The possibility of true muonium, bound states of $\mu^+\mu^-$, was realized~\cite{Marshak:1947zz} shortly after the leptonic nature of the muon was clarified~\cite{Lattes:1947mw,Lattes:1947mx,Lattes:1947my}. The positronium, the lightest quantum electrodynamics (QED) atom, which was discussed as early as the 1930s~\cite{1934AN-253-93M} and observed more than three-quarters of a century ago~\cite{Deutsch:1951zza}, leading to extensive studies. Muonium, the bound states of $\mu^+e^-$ was discovered earlier~\cite{Hughes:1960zz}, predating the first detailed discussions of true munion ($\mu^+\mu^-$)~\cite{Sov.J,Bull.Am.}. Although positronium, muonium~\cite{Hughes:1960zz}, $\pi\mu$ atoms~\cite{Coombes:1976hi} and dipositronium (the ($e^+e^-$)($e^+e^-$) molecule~\cite{Cassidy}) have been successfully produced and studied, the production of $\mu^+\mu^-$ bound states is exceptionally rare, even compared to the production of hadronic atoms. This feature poses a challenge in producing enough true muonium atoms for spectroscopic studies~\cite{Karshenboim:1998am}.
%the production of bound states of the $\mu^+\mu^-$ system is a rarer event than the process of hadronic atoms production and it is difficult to expect that enough atoms will be produced for spectroscopic studies~\cite{Karshenboim:1998am}.%{The cross section via two photon fustion is the magnitude of $\nu$b}
To date, the true muonium has not yet been observed experimentally~\cite{Dai:2024imb,Francener:2024eep}. However, fortunately, since true muonium is governed purely by QED, it allows for precise predictions and calculations.

Theoretically, the bound state of $\mu^+\mu^-$ can be produced in high-energy colliders, fixed target experiments and certain particle decays. Various proposals have been made to detect the true muonium in physical processes using high-energy experimental setups, such as $\pi^- p\to (\mu^+ \mu^-)n$~\cite{Bilenikii:1969}, $\gamma Z\to (\mu^+ \mu^-)Z$~\cite{Bilenikii:1969,Francener:2021wzx}, $e Z\to e(\mu^+ \mu^-)Z$~\cite{Arteaga-Romero:2000mwd,Krachkov:2017afm,Banburski:2012tk}, $Z_1 Z_2\to Z_1 Z_2(\mu^+\mu^-)$~\cite{Ginzburg:1998df,Chen:2012ci,Yu:2013uka,Azevedo:2019hqp,Yu:2022hdt,Francener:2021wzx} (where $Z$ indicates a heavy nucleus), $\mu^+\mu^-$ collisions~\cite{Hughes:1971}, $\eta\to (\mu^+\mu^-)\gamma$~\cite{Nemenov:1972,CidVidal:2019qub}, $e^+ e^-$ collisions~\cite{Moffat:1975uw,Brodsky:2009gx,Gargiulo:2023tci,Gargiulo:2024zyc}, $K_L\to (\mu^+\mu^-)\gamma$~\cite{Ji:2017lyh}, $B\to K^{(*)} (\mu^+\mu^-)$~\cite{Fael:2018ktm}, etc. 

%The collision process $e^+e^-\rightarrow(\mu^+\mu^-)$~\cite{Moffat:1975uw} was calculated following the discovery of $e^+e^-\rightarrow\mu^+\mu^-$. In Ref.~\cite{Brodsky:2009gx}, taken into account the energy resolution of the center of mass system and the indistinguishability of productions occurring below and above the $\mu^+\mu^-$ mass threshold~\cite{Fadin:1990wx}, the authors proposed two methods for detecting the true muonium in $e^+e^-$ colliders. The first method involves the process $e^+e^-\to (\mu^+\mu^-)$, using a asymmetric collider with beam merging a small crossing angle, which corresponds to the production of ortho-true muonium. The second one, $e^+e^-\to (\mu^+\mu^-) + \gamma$, leads to the production of both para- and ortho-true muonium. However, both detection methods face significant challenges due to the unprecedented QED backgrounds. In the search for para-true muonium, which predominantly decays into two photons, the dominant background comes from two photon process events ($e^+e^-\to \gamma\gamma + \gamma_{ISR}$). For ortho-true muonium, which mainly decays into $e^+e^-$, the primary background arises from the BhaBha sacttering events ($e^+e^-\to e^+e^- (+ \gamma_{ISR/FSR})$). \textbf{Here $\gamma_{ISR}$ and $\gamma_{FSR}$ represents the initial- and final-state radiative photon, respectively.} These QED backgrounds demand sophisticated techniques, such as greatly improved vertex resolution for the charged and neutral tracks or a exceptionally high integrated luminosity, to distinguish the signal events from background noise. 

The collision process $e^+e^-\rightarrow(\mu^+\mu^-)$~\cite{Moffat:1975uw} was calculated following the discovery of $e^+e^-\rightarrow\mu^+\mu^-$. In Ref.~\cite{Brodsky:2009gx}, considering the energy resolution of the center of mass system and the indistinguishability of productions occurring below and above the $\mu^+\mu^-$ mass threshold~\cite{Fadin:1990wx}, the authors proposed two methods for detecting the true muonium in $e^+e^-$ colliders. The first method involves the process $e^+e^-\to (\mu^+\mu^-)$, using an asymmetric collider with beam merging a small crossing angle, which corresponds to the production of ortho-true muonium. The second one, $e^+e^-\to (\mu^+\mu^-) + \gamma$, leads to the production of both para- and ortho-true muonium.  The first scheme requires the construction of a new collider~\cite{Bogomyagkov:2017uul},  while the second approach necessitates the accumulation of an enormous sample at lower energy point as the event rate is proportional to $1/s$~\cite{Brodsky:2009gx}. However, in practical experiments, it is generally infeasible to collect such a huge dataset at non-resonant energy points. In addition, both detection methods face significant challenges due to the unprecedented QED backgrounds. In the search for para-true muonium, which predominantly decays into two photons, the dominant background comes from two-photon process events ($e^+e^-\to \gamma\gamma + \gamma_{ISR}$). For ortho-true muonium, which mainly decays into $e^+e^-$, the primary background arises from the BhaBha scattering events ($e^+e^-\to e^+e^- (+ \gamma_{ISR/FSR})$). Here $\gamma_{ISR}$ and $\gamma_{FSR}$ represent the initial- and final-state radiative photon, respectively. These QED backgrounds demand sophisticated techniques, such as greatly improved vertex resolution for the charged and neutral tracks, to distinguish the signal events from background noise.

In this paper, we shall focus on the vector meson decays, specifically $J/\psi\rightarrow (\mu^+\mu^-) + \gamma$, to further explore potential detection strategies for true muonium production.
%It is similar to the process $e^+e^-\rightarrow (\mu^+\mu^-) + \gamma$ if we replace $e^+e^-$ by positronium, which means adding an evolving process from positronium to free electrons. 
This process is analogous to $e^+e^-\rightarrow (\mu^+\mu^-) + \gamma$, with the key difference being that $e^+e^-$ is replaced with $J/\psi$. It is important to note that using the $J/\psi$ sample directly taken at $\sqrt{s} = 3.097$~GeV, there will also encounter the significant QED backgrounds mentioned above. But when the vertex resolution is insufficient to discriminate the signal from QED backgrounds, we can attempt to utilize the $J/\psi$ data from the hadronic decay of $\psi(3686)$, such as $\psi(3686)\to \pi^+\pi^- J/\psi$, where the QED background could be removed well by applying a selection criterion on the recoiling mass of the $\pi^+\pi^-$ system~\cite{BESIII:2013csc,BESIII:2021ocn, BESIII:2012lxx}. It is important to note that the production cross section of the 
$\psi(3686)$  meson is roughly 20\% of the 
$J/\psi$, consequently, the available $J/\psi$ sample derived from 
$\psi(3686)$ decays—accounting for the branching fraction of 
$\psi(3686)\to \pi^+ \pi^- J/\psi$ and the reconstruction efficiency—will be approximately 0.04 times the size of the sample obtained at 3.097~GeV with the same integrated luminosity.

This paper is arranged in the following way. In Sec.~\ref{sec:amplitude}, the decay amplitude and the squared amplitude at small invariant mass of the lepton pair are derived. The phase space at small invariant mass is calculated in Sec.~\ref{sec:phasespace}. In Sec.~\ref{sec:width}, we estimate the decay width, incorporating both bound state contributions and Coulomb rescattering effects above the threshold. And in Sec.~\ref{sec:numerical}, we present the numerical evaluations of the branching ratio and discuss its experimental accessibility.

\section{The decay amplitude for $J/\psi \to \gamma \mu^+\mu^-$}\label{sec:amplitude}

\begin{figure}
    \centering
   \includegraphics[width=0.9\linewidth]{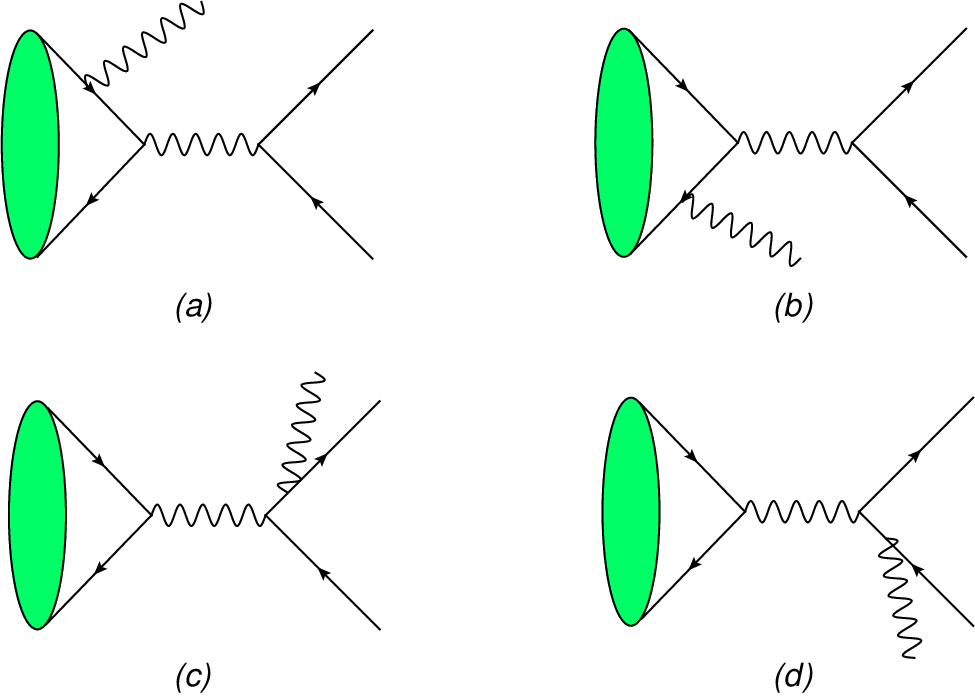}
    \caption{The Feynman diagrams for $J/\psi \to \gamma \mu^+\mu^-$ decay. Panels (a) and (b) do not contribute.}
    \label{fig:JToTM}
\end{figure}

%\subsection{Amplitude}
The Feynman diagrams for $J/\psi\to \gamma \mu^+\mu^-$ decay are shown in Fig.~\ref{fig:JToTM}. In the process, the photon can be radiated either from the $J/\psi$ meson or from outgoing muons ($\mu^+$ and $\mu^-$). However, due to the conservation of parity, the first two diagrams, where the photon is radiated directly from the $J/\psi$ meson, yield zero contribution, a result that can be rigorously verified through explicit calculations. Consequently, the nonzero contributions arise from final state radiation as depicted in Fig.~\ref{fig:JToTM} (c) and (d).

We treat the $J/\psi$ meson as a nonrelativistic system composed of a charm quark ($c$) and an anti-charm quark ($\bar c$). Under the nonrelativistic approximation, one has $P_{J/\psi}^2 = M_{J/\psi}^2\approx 4m_{c}^2$, where $M_{J/\psi}$ is the mass of the $J/\psi$ meson and $m_c$ the charm quark mass. The  momenta of the $\mu^-$, $\mu^+$ and the radiated photon $\gamma$ are denoted as $k_1$, $k_2$ and $k$, respectively. The invariant mass of the $\mu^+\mu^-$ system is denoted as $\sqrt{s_1}$. We focus on the threshold production of the lepton pair, where $s_1\approx 4m_{\mu}^2\ll 4m_c^2$. In the rest frame of $J/\psi$, the energy of the photon, $E_{\gamma}$, is nearly $E_{\gamma}\sim m_c$. In this scenario, the lepton pair is nearly collinear, and the photon and lepton pair are produced back-to-back, maintaining momentum conservation in the decay process.

To calculate the decay amplitude, we adopt the spin projector for the spin-triplet $c\bar c$ state as given in Ref.~\cite{Petrelli:1997ge},
\begin{align}
		\Pi_1 (P, \epsilon)= \frac{1}{\sqrt{8m_c^3}} \left(   \slashed p + m_c\right) \slashed  \epsilon\left(  \slashed  p-m_c\right),
\end{align}
where $p$ is the momentum of the $c$ or $\bar c$, respectively; $\epsilon$ is the polarization vector of the spin-triplet $c\bar c$ state. The normalization of the above projection operator corresponds to a relativistic normalization of the projected state. In nonrelativistic approximation, the momentum of $J/\psi$ is given by $P_{J/\psi}\approx 2p$.

The decay amplitude of $J/\psi\to \gamma \mu^+\mu^-$ is calculated as
\begin{align}
&	i \mathcal{M}(J/\psi\to \gamma \mu^+\mu^-)= \operatorname{Tr} \bigg[ \Pi_1 (P,\epsilon_{J/\psi})(-i e_Q e\gamma^{\rho}) \bigg] \frac{-i}{4m_c^2}\nonumber\\
 & \times\bar u (k_1)\bigg[(-i e\gamma_{\rho} )\frac{i}{-\slashed k-\slashed k_2- m_{\mu}}(-i e \slashed \epsilon^*) \nonumber\\
 &+ (-i e \slashed \epsilon^*) \frac{i}{\slashed k+\slashed k_1-m_{\mu}} (-i e \gamma_{\rho})\bigg]v(k_2) \sqrt{\frac{1}{4\pi}}R(0)\nonumber\\
	=& -i    e_Q e^3 \sqrt{2 m_c} \epsilon_{J/\psi}^{\rho} \frac{1}{2m_c^2} \sqrt{\frac{1}{4\pi}}R(0)\nonumber\\
 &\bar u (k_1)\bigg[ \gamma_{\rho} \frac{-\slashed k-\slashed k_2+ m_{\mu}}{2k\cdot k_2 } \slashed \epsilon^* +  \slashed \epsilon^* \frac{\slashed k+\slashed k_1+m_{\mu}}{ 2 k \cdot k_1}  \gamma_{\rho} \bigg]v(k_2) ,
\end{align}	
where $R(0)$ is the radial wave function of $J/\psi$ at the origin, $\epsilon_{J/\psi}$ and $\epsilon$ are the polarization vectors of $J/\psi$ and $\gamma$, respectively. $e$ represents the elementary electric charge, while $e_Q=2/3$ denotes the charge of the charm quark in units of $e$.

The squared amplitude, averaged over the $J/\psi$ polarization states, is given by
\begin{align}
&\frac13 \sum_{\mathrm{pols.}}	|i \mathcal{M}|^2 =   e_Q^2 e^6 \frac{1}{2 m_c^3}\frac{R^2(0)}{4\pi} \frac13 \bigg(-g^{\rho\rho'}+\frac{p^{\rho}p^{\rho'}}{m^2}\bigg) \nonumber\\
 &\times\operatorname{Tr} \bigg[\bigg(\gamma_{\rho} \frac{-\slashed k-\slashed k_2+ m_{\mu}}{2k\cdot k_2 }   \gamma_{\nu} +    \gamma_{\nu} \frac{\slashed k+\slashed k_1+m_{\mu}}{ 2 k \cdot k_1}  \gamma_{\rho} \bigg) \nonumber\\
 &\times(\slashed k_2-m_{\mu}) \bigg(\gamma^{\nu}\frac{-\slashed k-\slashed k_2+ m_{\mu}}{2k\cdot k_2 } \gamma_{\rho'}  +   \gamma_{\rho'} \frac{\slashed k+\slashed k_1+m_{\mu}}{ 2 k \cdot k_1}    \gamma^{\nu}\bigg) \nonumber\\
 &\times(\slashed k_1+ m_{\mu})\bigg]  ,
\end{align}	
where the factor $1/3$ comes from averaging over the three polarization states of the $J/\psi$.

In the near threshold production of the lepton pair, where the invariant mass $\sqrt{s_1}$ of the $\mu^+\mu^-$ pair is much smaller than the mass of $J/\psi$,  the above expression can be further simplified. When $s_1\ll 4m_c^2$, the squared amplitude can be approximated as
\begin{align}
\frac13 \sum_{\mathrm{pols.}}	|i \mathcal{M}|^2 \bigg\vert_{s_1\ll 4m_c^2}\approx	   \frac{e_Q^2 e^6}{3 \pi m_c^3}   \bigg(\frac{x_1}{x_2}+\frac{x_2}{x_1}\bigg) R^2(0),
\end{align}
where $x_{1,2}\equiv E_{1,2}/m_c$ are the energy fractions of the $\mu^-$ and $\mu^+$, with $E_{1,2}$ being their energies, respectively.

\section{The phase space integral at small invariant mass}\label{sec:phasespace}
Then we turn to the three-body phase space integral, which is defined as
\begin{align}
	\int &\frac{d^3 \vec{k}}{(2\pi)^3 2E_\gamma} \frac{d^3 \vec{k}_1}{(2\pi)^3 2E_1} \frac{d^3 \vec{k}_2}{(2\pi)^3 2E_2} \nonumber\\
 &\times (2\pi)^4 \delta^4 (2p-k-k_1-k_2),
\end{align}
where $\vec{k}_1, \vec{k}_2, \vec{k}$ are the three momenta of $\mu^-$, $\mu^+$ and the photon, respectively, and $E_1, E_2, E_\gamma$ are their corresponding energies. Integrating over the photon three momentum $\vec{k}$ and recalling the invariant mass of the muon pair $s_1 = (k_1 + k_2)^2$, one has
\begin{align}
&\frac{1}{(2\pi)^5}	\int  \frac{d^3 \vec{k}_1}{ 2E_1} \frac{d^3 \vec{k}_2}{ 2E_2}    \delta (4m_c^2+s_1 -4  m_c (E_1+E_2)).
\end{align}
With spherical coordinates, the expression becomes 
\begin{align}
	&\frac{1}{(2\pi)^5}	\int  \frac{|\vec{k}_1|^2 d |\vec{k}_1| d\cos\theta_1 d\phi_1}{ 2E_1} \frac{|\vec{k}_2|^2 d |\vec{k}_2| d\cos\theta_{12}d\phi_2}{ 2E_2}  \nonumber\\
 &\times\delta (4m_c^2+s_1 -4  m_c (E_1+E_2)),
\end{align}
where $\theta_{12}$ is the angle between $\vec{k}_1$ and $\vec{k}_2$. The integrals over $\phi_1$ and $\phi_2$ can be calculated directly. Now we convert the integrals over $|\vec{k}_1|$ and $|\vec{k}_2|$ into the integral over $E_1$ and $E_2$ using the relations: 
\begin{align}
d E_1=	\frac{|\vec{k}_1|}{E_1} d |\vec{k}_1| ,  ~~~~ dE_2=\frac{|\vec{k}_2|}{E_2} d |\vec{k}_2|.
\end{align}
The phase space integral then becomes
\begin{align}
	&\frac{1}{(2\pi)^3}	\frac{1}{4}\int  d E_1 d\cos\theta_1  d E_2 d\cos\theta_{12}  |\vec{k}_1| |\vec{k}_2| \nonumber\\
 &\times\delta (4m_c^2+s_1 -4  m_c (E_1+E_2)).
\end{align}
On the other hand, on account of 
\begin{align}
	s_1=(k_1+k_2)^2 =2m_{\mu}^2+2 (E_1 E_2- |\vec{k}_1| |\vec{k}_2| \cos \theta_{12}),
\end{align}
we can convert the integral over $\cos\theta_{12}$ into an integral over $s_1$,
\begin{align}
	\int_{-1}^1 d\cos \theta_{12} =\int_{s_{\mathrm{min}}}^{s_{\mathrm{max}}}\frac{1}{2|\vec{k}_1| |\vec{k}_2| } ds_1,
\end{align}
where the limits are given by:
\begin{align}
	&s_{\mathrm{min}}=2m_{\mu}^2+2 (E_1 E_2-\sqrt{E_1^2-m_{\mu}^2}\sqrt{E_2^2-m_{\mu}^2}) ,\\
	&s_{\mathrm{max}}=2m_{\mu}^2+2 (E_1 E_2+\sqrt{E_1^2-m_{\mu}^2}\sqrt{E_2^2-m_{\mu}^2}) .
\end{align}
These limits define the kinematically allowed range for $s_1$, completing the setup for the phase space integration.

After exchanging the order of integrals, one has
\begin{align}
	&\frac{1}{(2\pi)^3}	\frac{1}{2^3}\int_{-1}^1 d\cos\theta_1\iint  d E_1   d E_2 \int_{s_{\mathrm{min}}}^{s_{\mathrm{max}}}d s_1  \nonumber\\
&\times\delta (4m_c^2+s_1 -4  m_c (E_1+E_2))\nonumber\\
=	&\frac{1}{(2\pi)^3}	\frac{1}{2^3}\int_{-1}^1 d\cos\theta_1   \int ds_1 \iint_S d E_1  d E_2  \nonumber\\
&\times \delta (4m_c^2+s_1 -4  m_c (E_1+E_2)) ,
\end{align}
where $S$ is the region in $E_1-E_2$ plane with the boundaries given by the equation
\begin{align}
	s_{\mathrm{max}}(E_1,E_2)=s_1 ~~\mathrm{or}~~	s_{\mathrm{min}}(E_1,E_2)=s_1.
\end{align}
The range of $E_1$ and $E_2$ is given as below. The delta function constraint $4m_c^2+s_1 -4  m_c (E_1+E_2) = 0$ implies that the integral is nonzero only within this region. The boundary conditions are given by
\begin{align}
2m_{\mu}^2+2 \left(E_1 E_2-\sqrt{E_1^2-m_{\mu}^2}\sqrt{E_2^2-m_{\mu}^2} \right) = &s_1,\\
4m_c^2+s_1 -4  m_c (E_1+E_2)=&0.
\end{align}
Solving the two equations, we find
\begin{align}
E_1=&\frac{4m_c^3 s_1+m_c s_1^2 \pm \sqrt{s_1 (s_1-4m_{\mu}^2)m_c^2(s_1-4m_c^2)^2}}{8m_c^2 s_1},\\
E_2=&\frac{4m_c^3 s_1+m_c s_1^2 \mp \sqrt{s_1 (s_1-4m_{\mu}^2)m_c^2(s_1-4m_c^2)^2}}{8m_c^2 s_1}.
\end{align}
When $s_1\ll m_c^2$, these can be simplified to
\begin{align}
	E_1&= \frac{m_c}{2} \left(1\pm \sqrt{1-\frac{4m_{\mu}^2}{s_1}}\right), \\
 E_2&=\frac{m_c}{2} \left(1\mp \sqrt{1-\frac{4m_{\mu}^2}{s_1}}\right).
\end{align}
The velocity of the muon in the center-of-mass frame of the muon pair is $\beta=\sqrt{1-\frac{4m_{\mu}^2}{s_1}}$. Thus the final expression for the integration reads
\begin{align}
	&\frac{1}{(2\pi)^3}	\frac{1}{16}   \int  ds_1 \int_{\frac{1}{2}(1-\beta)}^{\frac{1}{2}(1+\beta)} d x_1   .
\end{align}
Here, $x_{1} = \frac{E_1}{m_c}$, as mentioned in earlier section, represents the energy fraction of the $\mu^-$ in the center-of-mass (c.o.m.) frame of $J/\psi$ meson, with the integration limits determined by the velocity $\beta$ of the muon.

\section{Coulomb resummation and Decay width}\label{sec:width}
With the amplitude and the phase space integral calculated above, the decay width for the process $J/\psi\to \gamma\mu^+\mu^-$ at small invariant mass of lepton pair reads
\begin{align}
	&\frac{d\Gamma(J/\psi\to \gamma \mu^+\mu^-)}{ds_1} \bigg\vert_{s_1\ll 4m_c^2}\nonumber\\
 = 
	&   R^2(0)  \frac{e_Q^2 \alpha^3 }{ 24\pi m_c^4}  \int_{\frac12(1-\beta)}^{\frac12(1+\beta)} dx_1   \bigg(\frac{x_1}{x_2}+\frac{x_2}{x_1}\bigg)\nonumber\\
 =&\frac{e_Q^2 \alpha^3 }{12\pi m_c^4}R^2(0)   \bigg( \ln\frac{1+\beta}{1-\beta}-\beta\bigg), 
\end{align}
where $\alpha=e^2/(4\pi)$ is the fine structure constant. In the last line we have adopted the approximation $x_2= 1-x_1+\frac{s_1}{4m_c^2}\approx 1-x_1$ and performed the $x_1$-integral. In the threshold limit $\beta\to 0$, the expression $\ln\frac{1+\beta}{1-\beta}-\beta\sim \beta$, causing the decay width to vanish as $\beta=0$. The above formula gives the differential decay width for the decay $J/\psi$ into a photon and a $\mu^+\mu^-$ pair with small invariant mass compared to the mass of $J/\psi$, but the lepton pair is not necessary bound each other. 

In the threshold region, the $\mu^+/\mu^-$ moves non-relativistically and can exchange Coulomb photons in the c.o.m. frame of the lepton pair. The effect of Coulomb photon exchange is of the order $\alpha/\beta\sim\mathcal{O}(1)$, which calls for a resummation to improve the convergence of the perturbative expansions. Such a resummation can be performed by solving the Green function of Schr\"{o}dinger equation with a Coulomb potential $-\alpha/r$. The impact of Coulomb rescattering of the non-relativistic particles in QED was first
discussed by Sommerfeld, Sakharov and Schwinger~\cite{Sommerfeld:1931qaf,Sakharov:1948plh,Schwinger:1973rv}. Analog to Ref.~\cite{Fadin:1990wx}, the decay width can be related to the imaginary part of the Coulomb Green function with
\begin{align}
&	\frac{d\Gamma(J/\psi\to \gamma(\mu^+\mu^-))}{ds_1} \nonumber\\
 =& \frac{d\Gamma(J/\psi\to \gamma \mu^+\mu^-) }{ds_1} \bigg\vert_{s_1\ll 4m_c^2}\frac{4\pi}{m_{\mu}^2 \beta} \operatorname{Im} G_{E+i\Gamma} (0,0), 
\end{align}
where $G_{E+i\Gamma}(\vec{r_1},\vec{r_2})$ denotes the Coulomb Green function, $E=\sqrt{s_1}-2m_{\mu}$ is the difference of the invariant mass of $(\mu^+\mu^-)$ and the threshold, and $\Gamma$ is the decay width of the lepton. The imaginary part of the Coulomb Green function is given by~\cite{Fadin:1987wz}
\begin{align}
&\operatorname{Im}	G_{E+i \Gamma}(0,0)\nonumber\\
=& \frac{m_{\mu}^2}{4\pi}\bigg[\frac{p_2}{m_{\mu}}+\frac{2p_s}{m_{\mu}}\arctan \frac{p_2}{p_1} +\frac{2p_s^2}{m_{\mu}^2}\nonumber\\
&\times \sum_{n=1}^{\infty} \frac{1}{n^4} \frac{\Gamma p_s n+p_2 (n^2 \sqrt{E^2+\Gamma^2}+p_s^2/m_{\mu})}{\left(E+\frac{p_s^2}{m_{\mu} n^2}\right)^2+\Gamma^2}\bigg],
\end{align}
where 
\begin{align}
	p_s=\frac12 m_{\mu} \alpha, ~~ p_{1,2} = \sqrt{\frac{m_{\mu}}{2}(\sqrt{E^2+\Gamma^2}\mp E)}.
\end{align}
Because the life time of muon is about 2~$\mu$s, significantly longer than the life time of true muonium, which is the order of several picoseconds, we treat muon as a stable particle and take the narrow width limit, setting $\Gamma = 0$ setting for most practical calculations involving true muonium. However, when analyzing the production of true tauonium ($\tau^+\tau^-$), the finite decay width $\Gamma$ of the tau lepton must be taken into account, as it has a substantial impact on the system's dynamics and observable effects.

By changing the variable from $s_1 $ to $E$ with $E=\sqrt{s_1}-2m_{\mu}$, one has
\begin{align}
&\frac{d\Gamma(J/\psi\to \gamma (\mu^+\mu^-))}{d E}\nonumber\\
=&\frac{e_Q^2 \alpha^3 }{12\pi m_c^4}R^2(0)  2(E+2m_{\mu})   \left( \ln\frac{1+\beta}{1-\beta}-\beta\right) \nonumber\\
&\times\frac{4\pi}{m_{\mu}^2 \beta}\operatorname{Im} G_{E+i0} (0,0) .
\end{align}
To get rid of the radial wave function at the origin, which is a nonperturbative quantity, we introduce $R$ as the ratio of the decay widths of $J/\psi\to \gamma (\mu^+\mu^-)$ and $J/\psi\to \mu^+\mu^-$,
\begin{align}
    R=\frac{\Gamma(J/\psi\to \gamma (\mu^+\mu^-))}{\Gamma(J/\psi\to \mu^+\mu^-)} ,
\end{align}
where the decay width of $J/\psi\to \mu^+\mu^-$ is calculated as
\begin{align}
	\Gamma(J/\psi \to \mu^+ \mu^-)  
			=       \frac{   \alpha^2 e_Q^2}{3 m_c^2} \sqrt{1-\frac{m_{\mu}^2}{m_c^2}}   \left(1+\frac{m_{\mu}^2}{2m_c^2}\right)  R^2 (0) .
\end{align}

For bound states, which are below the threshold with a binding energy $E<0$, we finds that
\begin{align}
&	\frac{d\Gamma(J/\psi\to \gamma (\mu^+\mu^-))}{d E}\bigg\vert_{E<0}\nonumber\\
 =&\frac{e_Q^2 \alpha^3 }{12\pi m_c^4}R^2(0)  2(E+2m_{\mu})  \left( \ln\frac{1+\beta}{1-\beta}-\beta\right) \nonumber\\
 &\times\frac{4\pi}{m_{\mu}^2 \beta}\sum_{n=1}^{\infty} \frac{\alpha^3 m_{\mu}^3 }{8n^3} \delta \left(E+\frac{\alpha^2 m_{\mu}}{4n^2}\right).
\end{align}
The argument in the delta function can be identified as $E-E_n$, where $E_n=-\alpha^2 m_{\mu}/(4n^2)$ represents the binding energy of the state with principal quantum number $n$. The squared wave function at the origin for these bound states is given by 
\begin{align}
    |\psi_n (0)|^2 =\frac{\alpha^3 m_{\mu}^3}{8 \pi n^3}.
\end{align}
Integrating over $E$ yields the total decay width. When this result is divided by the decay width $\Gamma(J/\psi\to \mu^+\mu^-)$, one has
 \begin{align}
     R\vert_{E<0}\approx &\frac{\alpha^4 m_{\mu}^2}{\sqrt{1-\frac{m_{\mu}^2}{m_c^2}}(m_{\mu}^2+2m_c^2)}\zeta(3)\nonumber\\
     \approx & \frac{\alpha^4 m_{\mu}^2}{ 2m_c^2}\zeta(3), 
 \end{align}
where $\zeta(s)$ is the Riemann zeta function. We also verify the above result in another approach, with the help of spin-projector and the wave function of true para-muonium at the origin. This method provides an independent cross-check of our calculation.

When $E>0$, meaning the lepton pair is above the threshold, one has
\begin{align}
&\frac{d\Gamma(J/\psi\to \gamma (\mu^+\mu^-))}{d E} \bigg|_{E>0}\nonumber\\
=&\frac{e_Q^2 \alpha^3 }{12\pi m_c^4}R^2(0)   2(E+2m_{\mu})   \left( \ln\frac{1+\beta}{1-\beta}-\beta\right) \nonumber\\
&\times\frac{4\pi}{m_{\mu}^2 \beta} \frac{\alpha m_{\mu}^2 }{4\left(1-e^{-\frac{\alpha \pi m_{\mu}}{\sqrt{E m_{\mu}}}}\right)}.
\end{align}
One can identify that the term in the last line is the Sommerfeld-Schwinger-Sakharov factor
\begin{align}
    S(\beta)\equiv \frac{\pi \alpha/\beta}{1-e^{-\frac{\alpha \pi m_{\mu}}{\sqrt{E m_{\mu}}}}},
\end{align}
which accounts for the Coulomb enhancement of the decay rate near the threshold. 
For the ratio $R$ of decay widths, one has
\begin{align}
    R\vert_{E>0}\approx & \frac{\alpha}{\pi}\int_0^{\Lambda}  dE (E+2m_{\mu})\frac{      ( \ln\frac{1+\beta}{1-\beta}-\beta) S(\beta)} {     \sqrt{1-\frac{m_{\mu}^2}{m_Q^2}}   (2m_Q^2+m_{\mu}^2)   }\nonumber\\
    \approx & \alpha^2 \frac{m_{\mu}}{m_c^2}\int_0^{\Lambda} \frac{ dE }{1-e^{-\pi \alpha\sqrt{\frac{m_{\mu}}{E}}}},
\end{align}
where the integral is calculated in an energy window $[0, \Lambda]$.

\section{ Numerical results}\label{sec:numerical}
With the formulas derived in the previous sections, we now present numerical results for the decay process $J/\psi\to \gamma (\mu^+\mu^-)$. Following parameter values adopted from the Particle Data Group (PDG)~\cite{ParticleDataGroup:2024cfk} are used in the numerical calculation,
\begin{align}
  m_{\mu}=105.66~\mathrm{MeV},~~m_c=1.27~\mathrm{GeV},~~\alpha=\frac{1}
  {137}.
\end{align}
With these inputs, we get the ratio $R$ for the bound state contributions
\begin{align}
  R\vert_{E<0}\approx 1.18 \times 10^{-11}.
\end{align}
Using the branching fraction of $J/\psi\to \mu^+\mu^-$~\cite{ParticleDataGroup:2024cfk} as follows:
\begin{align}
  \mathrm{Br}(J/\psi\to \mu^+\mu^-) = 5.961\%,
\end{align}
we can then calculate the branching fraction of $J/\psi\to \gamma (\mu^+\mu^-)$. The result reads
\begin{align}
  \mathrm{Br}(J/\psi \to \gamma (\mu^+\mu^-))\approx 7.03\times 10^{-13}.
\end{align}
%BESIII has accumulated approximately $2.7\times10^{9}$~$\psi(3686)$ events~\cite{BESIII:2024lks}, which suggests that detecting true para-muonium via $J/\psi$ radiative decay remains infeasible with current data. However, at the future super tau-charm facility (STCF)~\cite{Achasov:2023gey}, an anticipated annual production of $3.4\times 10^{12}$~$J/\psi$ events would result in $2\sim 3$ observable $J/\psi\to \gamma (\mu^+\mu^-)$ events per year. While detection at STCF presents a challenge, the facility’s increased event rate offers promising prospects for this rare process search.
BESIII has accumulated approximately $10^{10}$~$J/\psi$ events~\cite{BESIII:2021cxx} and $2.7\times10^{9}$~$\psi(3686)$ events~\cite{BESIII:2024lks}, which suggests that detecting true para-muonium via $J/\psi$ radiative decay remains infeasible with current data. However, at the future super tau-charm facility (STCF)~\cite{Achasov:2023gey}, an anticipated annual production of $3.4\times 10^{12}$~$J/\psi$ events would result in $2\sim 3$ observable $J/\psi\to \gamma (\mu^+\mu^-)$ events per year. While detection at STCF presents a challenge, the facility’s increased event rate offers promising prospects for this rare process search.

Considering the resolution limitations in the experimental reconstruction, an uncertainty exists for the invariant mass of the $(\mu^+\mu^-)$ pair, as well as $E = \sqrt{s_1} - 2m_{\mu}$. As discussed in Ref.~\cite{Brodsky:2009gx}, the $\mu^+\mu^-$ pair below and just above the threshold could be indistinguishable due to the uncertainty. To identify the true muonium from the other events, the resolution should be as small as the width of the energy spectrum of the true muonium. If the energy window is taken as the energy difference between the ground state and the threshold, namely, $\Lambda = -E_0 = \alpha^2 m_{\mu}/4$, as what has been done in Ref.~\cite{Brodsky:2009gx}, one gets
\begin{align}
  R\vert_{E>0}\approx 4.91 \times 10^{-12},
\end{align}
which is comparable in order of magnitude to the bound-state contribution. We note that this value could increase quickly if the energy window is broadened. The true para-muonium should be reconstructed from its $\gamma\gamma$-decay mode in experiments. The measured precision of the true-muonium invariant mass is determined by the resolution of the photon energy, which is around several MeV at the BESIII~\cite{BESIII:2009fln}. If $\Lambda\sim$ MeV, $R|_{E>0}$ can reach as high as $10^{-8}$, indicating that thousands of events per year could be collected at the STCF. Therefore, better energy resolution is crucial to identify the real true-muonium events.

True muonium, once produced, can be identified by its characteristic decay length. The lifetime of ground-state true para-muonium is approximately 0.6 ps. In the lab frame, true muonium generated via radiative $J/\psi$ decay is boosted with a Lorentz factor of $\gamma\approx 7.4$, 
%implying that the decay vertex of true muonium will be displaced from the $J/\psi$ decay point by about 1 mm. This separation falls within the detector’s path resolution capabilities.
leading to an average travel distance of around 1.3~mm in the detector. The distance allows the decay vertex of true muonium to be spatially separated from that of $J/\psi$. Therefore this separation provides a distinct advantage for distinguishing true muonium from background processes, such as two photon process $e^+e^-\to \gamma\gamma + \gamma_{ISR}$ and $J/\psi \to \gamma P \to 3\gamma$ ($P$ denotes the pseudo-scalar mesons like $\pi^0$, $\eta$, $\eta^\prime$ and $\eta_c$), all of which have zero flight distance within the detector. If the detector vertex resolution is able to reach 0.4~mm, this separation is well within measurable limits, allowing for the identification of true muonium. We hope that with the continuous accumulation of data and advancements in the neutral track vertex reconstruction, enabling precise reconstruction of the true muonium decay length to effectively discriminate against backgrounds, observing genuine para-muonium in $J/\psi$ radiative decays will become feasible at the STCF.

\section{ Conclusion} \label{sec:conclusion}

In this paper, we have explored the potential for detecting true para-muonium through the radiative decay of the $J/\psi$ meson at electron-positron colliders. True muonium, a bound state of a muon and antimuon, remains an elusive particle despite decades of theoretical interest. The $J/\psi$ radiative decay, e.g.,
$J/\psi\to \gamma(\mu^+\mu^-)$, offers a unique and clean channel to detect the true muonium. The production of true para-muonium via this decay is primarily governed by QED processes. We presented a detailed analysis of the relevant Feynman diagrams and computed the decay width, accounting for the Sommerfeld enhancement due to the Coulomb interaction between the muon-antimuon pair. We demonstrated that the branching ratio of this channel is of the order $10^{-13}$, which makes
the true muonium difficult to be observed via $J/\psi$ decays in current experiments; however,
future high-energy collider experiments, such as the proposed Super Tau-Charm Facility, offer promising avenues for discovering this exotic QED bound state.

\section*{Acknowledgments}

The work of J.~P.~D. was supported by the National Natural Science Foundation of China (Grants No.~12165022) and Yunnan Fundamental Research Project under Contract No.~202301AT070162. The work of H.~B.~L. was supported by National Key R$\&$D Program of China under Contract No.~2023YFA1606000, National Natural Science Foundation of China under Contract No.~11935018. The work of S.~Z. was supported in part by the National Natural Science Foundation of China under Contract No.~12475098.

\bibliographystyle{apsrev}
\bibliography{JPsiToTM}% Produces the bibliography via BibTeX.

\begin{thebibliography}{48}
\expandafter\ifx\csname natexlab\endcsname\relax\def\natexlab#1{#1}\fi
\expandafter\ifx\csname bibnamefont\endcsname\relax
  \def\bibnamefont#1{#1}\fi
\expandafter\ifx\csname bibfnamefont\endcsname\relax
  \def\bibfnamefont#1{#1}\fi
\expandafter\ifx\csname citenamefont\endcsname\relax
  \def\citenamefont#1{#1}\fi
\expandafter\ifx\csname url\endcsname\relax
  \def\url#1{\texttt{#1}}\fi
\expandafter\ifx\csname urlprefix\endcsname\relax\def\urlprefix{URL }\fi
\providecommand{\bibinfo}[2]{#2}
\providecommand{\eprint}[2][]{\url{#2}}

\bibitem[{\citenamefont{Marshak and Bethe}(1947)}]{Marshak:1947zz}
\bibinfo{author}{\bibfnamefont{R.~E.} \bibnamefont{Marshak}} \bibnamefont{and}
  \bibinfo{author}{\bibfnamefont{H.~A.} \bibnamefont{Bethe}},
  \bibinfo{journal}{Phys. Rev.} \textbf{\bibinfo{volume}{72}},
  \bibinfo{pages}{506} (\bibinfo{year}{1947}).

\bibitem[{\citenamefont{Lattes et~al.}(1947{\natexlab{a}})\citenamefont{Lattes,
  Muirhead, Occhialini, and Powell}}]{Lattes:1947mw}
\bibinfo{author}{\bibfnamefont{C.~M.~G.} \bibnamefont{Lattes}},
  \bibinfo{author}{\bibfnamefont{H.}~\bibnamefont{Muirhead}},
  \bibinfo{author}{\bibfnamefont{G.~P.~S.} \bibnamefont{Occhialini}},
  \bibnamefont{and} \bibinfo{author}{\bibfnamefont{C.~F.}
  \bibnamefont{Powell}}, \bibinfo{journal}{Nature}
  \textbf{\bibinfo{volume}{159}}, \bibinfo{pages}{694}
  (\bibinfo{year}{1947}{\natexlab{a}}).

\bibitem[{\citenamefont{Lattes et~al.}(1947{\natexlab{b}})\citenamefont{Lattes,
  Occhialini, and Powell}}]{Lattes:1947mx}
\bibinfo{author}{\bibfnamefont{C.~M.~G.} \bibnamefont{Lattes}},
  \bibinfo{author}{\bibfnamefont{G.~P.~S.} \bibnamefont{Occhialini}},
  \bibnamefont{and} \bibinfo{author}{\bibfnamefont{C.~F.}
  \bibnamefont{Powell}}, \bibinfo{journal}{Nature}
  \textbf{\bibinfo{volume}{160}}, \bibinfo{pages}{453}
  (\bibinfo{year}{1947}{\natexlab{b}}).

\bibitem[{\citenamefont{Lattes et~al.}(1947{\natexlab{c}})\citenamefont{Lattes,
  Occhialini, and Powell}}]{Lattes:1947my}
\bibinfo{author}{\bibfnamefont{C.~M.~G.} \bibnamefont{Lattes}},
  \bibinfo{author}{\bibfnamefont{G.~P.~S.} \bibnamefont{Occhialini}},
  \bibnamefont{and} \bibinfo{author}{\bibfnamefont{C.~F.}
  \bibnamefont{Powell}}, \bibinfo{journal}{Nature}
  \textbf{\bibinfo{volume}{160}}, \bibinfo{pages}{486}
  (\bibinfo{year}{1947}{\natexlab{c}}).

\bibitem[{\citenamefont{{Mohorovi{\v{c}}i{\'c}}}(1934)}]{1934AN-253-93M}
\bibinfo{author}{\bibfnamefont{S.}~\bibnamefont{{Mohorovi{\v{c}}i{\'c}}}},
  \bibinfo{journal}{Astronomische Nachrichten} \textbf{\bibinfo{volume}{253}},
  \bibinfo{pages}{93} (\bibinfo{year}{1934}).

\bibitem[{\citenamefont{Deutsch}(1951)}]{Deutsch:1951zza}
\bibinfo{author}{\bibfnamefont{M.}~\bibnamefont{Deutsch}},
  \bibinfo{journal}{Phys. Rev.} \textbf{\bibinfo{volume}{82}},
  \bibinfo{pages}{455} (\bibinfo{year}{1951}).

\bibitem[{\citenamefont{Hughes et~al.}(1960)\citenamefont{Hughes, McColm,
  Ziock, and Prepost}}]{Hughes:1960zz}
\bibinfo{author}{\bibfnamefont{V.~W.} \bibnamefont{Hughes}},
  \bibinfo{author}{\bibfnamefont{D.~W.} \bibnamefont{McColm}},
  \bibinfo{author}{\bibfnamefont{K.}~\bibnamefont{Ziock}}, \bibnamefont{and}
  \bibinfo{author}{\bibfnamefont{R.}~\bibnamefont{Prepost}},
  \bibinfo{journal}{Phys. Rev. Lett.} \textbf{\bibinfo{volume}{5}},
  \bibinfo{pages}{63} (\bibinfo{year}{1960}).

\bibitem[{\citenamefont{S.~Bilen’kii and Tkebuchava}(1969)}]{Sov.J}
\bibinfo{author}{\bibfnamefont{L.~N.} \bibnamefont{S.~Bilen’kii},
  \bibfnamefont{N.~van~Hieu}} \bibnamefont{and}
  \bibinfo{author}{\bibfnamefont{F.}~\bibnamefont{Tkebuchava}},
  \bibinfo{journal}{Sov.J} \textbf{\bibinfo{volume}{10}}, \bibinfo{pages}{469}
  (\bibinfo{year}{1969}).

\bibitem[{\citenamefont{Hughes and B.~Maglic}(1971)}]{Bull.Am.}
\bibinfo{author}{\bibfnamefont{V.}~\bibnamefont{Hughes}} \bibnamefont{and}
  \bibinfo{author}{\bibfnamefont{B.}~\bibnamefont{B.~Maglic}},
  \bibinfo{journal}{Am. Phys. Soc.} \textbf{\bibinfo{volume}{16}},
  \bibinfo{pages}{69} (\bibinfo{year}{1971}).

\bibitem[{\citenamefont{Coombes et~al.}(1976)}]{Coombes:1976hi}
\bibinfo{author}{\bibfnamefont{R.}~\bibnamefont{Coombes}} \bibnamefont{et~al.},
  \bibinfo{journal}{Phys. Rev. Lett.} \textbf{\bibinfo{volume}{37}},
  \bibinfo{pages}{249} (\bibinfo{year}{1976}).

\bibitem[{\citenamefont{Cassidy~DB}(2007)}]{Cassidy}
\bibinfo{author}{\bibfnamefont{M.~A.~J.} \bibnamefont{Cassidy~DB}},
  \bibinfo{journal}{Nature} \textbf{\bibinfo{volume}{7159}},
  \bibinfo{pages}{195} (\bibinfo{year}{2007}).

\bibitem[{\citenamefont{Karshenboim et~al.}(1998)\citenamefont{Karshenboim,
  Ivanov, Jentschura, and Soff}}]{Karshenboim:1998am}
\bibinfo{author}{\bibfnamefont{S.~G.} \bibnamefont{Karshenboim}},
  \bibinfo{author}{\bibfnamefont{V.~G.} \bibnamefont{Ivanov}},
  \bibinfo{author}{\bibfnamefont{U.~D.} \bibnamefont{Jentschura}},
  \bibnamefont{and} \bibinfo{author}{\bibfnamefont{G.}~\bibnamefont{Soff}},
  \bibinfo{journal}{J. Exp. Theor. Phys.} \textbf{\bibinfo{volume}{86}},
  \bibinfo{pages}{226} (\bibinfo{year}{1998}).

\bibitem[{\citenamefont{Dai and Zhao}(2024)}]{Dai:2024imb}
\bibinfo{author}{\bibfnamefont{J.-P.} \bibnamefont{Dai}} \bibnamefont{and}
  \bibinfo{author}{\bibfnamefont{S.}~\bibnamefont{Zhao}},
  \bibinfo{journal}{Phys. Rev. D} \textbf{\bibinfo{volume}{109}},
  \bibinfo{pages}{054022} (\bibinfo{year}{2024}), \eprint{2401.04681}.

\bibitem[{\citenamefont{Francener et~al.}(2024)\citenamefont{Francener,
  Goncalves, Moreira, and Santos}}]{Francener:2024eep}
\bibinfo{author}{\bibfnamefont{R.}~\bibnamefont{Francener}},
  \bibinfo{author}{\bibfnamefont{V.~P.} \bibnamefont{Goncalves}},
  \bibinfo{author}{\bibfnamefont{B.~D.} \bibnamefont{Moreira}},
  \bibnamefont{and} \bibinfo{author}{\bibfnamefont{K.~A.}
  \bibnamefont{Santos}}, \bibinfo{journal}{Phys. Lett. B}
  \textbf{\bibinfo{volume}{854}}, \bibinfo{pages}{138753}
  (\bibinfo{year}{2024}), \eprint{2404.11610}.

\bibitem[{\citenamefont{S.~Bilenikii and Tkebuchava}(1969)}]{Bilenikii:1969}
\bibinfo{author}{\bibfnamefont{L.~N.} \bibnamefont{S.~Bilenikii},
  \bibfnamefont{N.~van~Hieu}} \bibnamefont{and}
  \bibinfo{author}{\bibfnamefont{F.}~\bibnamefont{Tkebuchava}},
  \bibinfo{journal}{Sov.J.Nucl.Phys.} \textbf{\bibinfo{volume}{10}},
  \bibinfo{pages}{469} (\bibinfo{year}{1969}).

\bibitem[{\citenamefont{Francener et~al.}(2022)\citenamefont{Francener,
  Goncalves, and Moreira}}]{Francener:2021wzx}
\bibinfo{author}{\bibfnamefont{R.}~\bibnamefont{Francener}},
  \bibinfo{author}{\bibfnamefont{V.~P.} \bibnamefont{Goncalves}},
  \bibnamefont{and} \bibinfo{author}{\bibfnamefont{B.~D.}
  \bibnamefont{Moreira}}, \bibinfo{journal}{Eur. Phys. J. A}
  \textbf{\bibinfo{volume}{58}}, \bibinfo{pages}{35} (\bibinfo{year}{2022}),
  \eprint{2110.03466}.

\bibitem[{\citenamefont{Arteaga-Romero
  et~al.}(2000)\citenamefont{Arteaga-Romero, Carimalo, and
  Serbo}}]{Arteaga-Romero:2000mwd}
\bibinfo{author}{\bibfnamefont{N.}~\bibnamefont{Arteaga-Romero}},
  \bibinfo{author}{\bibfnamefont{C.}~\bibnamefont{Carimalo}}, \bibnamefont{and}
  \bibinfo{author}{\bibfnamefont{V.~G.} \bibnamefont{Serbo}},
  \bibinfo{journal}{Phys. Rev. A} \textbf{\bibinfo{volume}{62}},
  \bibinfo{pages}{032501} (\bibinfo{year}{2000}), \eprint{hep-ph/0001278}.

\bibitem[{\citenamefont{Krachkov and Milstein}(2018)}]{Krachkov:2017afm}
\bibinfo{author}{\bibfnamefont{P.~A.} \bibnamefont{Krachkov}} \bibnamefont{and}
  \bibinfo{author}{\bibfnamefont{A.~I.} \bibnamefont{Milstein}},
  \bibinfo{journal}{Nucl. Phys. A} \textbf{\bibinfo{volume}{971}},
  \bibinfo{pages}{71} (\bibinfo{year}{2018}), \eprint{1712.09770}.

\bibitem[{\citenamefont{Banburski and Schuster}(2012)}]{Banburski:2012tk}
\bibinfo{author}{\bibfnamefont{A.}~\bibnamefont{Banburski}} \bibnamefont{and}
  \bibinfo{author}{\bibfnamefont{P.}~\bibnamefont{Schuster}},
  \bibinfo{journal}{Phys. Rev. D} \textbf{\bibinfo{volume}{86}},
  \bibinfo{pages}{093007} (\bibinfo{year}{2012}), \eprint{1206.3961}.

\bibitem[{\citenamefont{Ginzburg et~al.}(1998)\citenamefont{Ginzburg,
  Jentschura, Karshenboim, Krauss, Serbo, and Soff}}]{Ginzburg:1998df}
\bibinfo{author}{\bibfnamefont{I.~F.} \bibnamefont{Ginzburg}},
  \bibinfo{author}{\bibfnamefont{U.~D.} \bibnamefont{Jentschura}},
  \bibinfo{author}{\bibfnamefont{S.~G.} \bibnamefont{Karshenboim}},
  \bibinfo{author}{\bibfnamefont{F.}~\bibnamefont{Krauss}},
  \bibinfo{author}{\bibfnamefont{V.~G.} \bibnamefont{Serbo}}, \bibnamefont{and}
  \bibinfo{author}{\bibfnamefont{G.}~\bibnamefont{Soff}},
  \bibinfo{journal}{Phys. Rev. C} \textbf{\bibinfo{volume}{58}},
  \bibinfo{pages}{3565} (\bibinfo{year}{1998}), \eprint{hep-ph/9805375}.

\bibitem[{\citenamefont{Chen and Zhuang}(2012)}]{Chen:2012ci}
\bibinfo{author}{\bibfnamefont{Y.}~\bibnamefont{Chen}} \bibnamefont{and}
  \bibinfo{author}{\bibfnamefont{P.}~\bibnamefont{Zhuang}}
  (\bibinfo{year}{2012}), \eprint{1204.4389}.

\bibitem[{\citenamefont{Yu and Li}(2013)}]{Yu:2013uka}
\bibinfo{author}{\bibfnamefont{G.-M.} \bibnamefont{Yu}} \bibnamefont{and}
  \bibinfo{author}{\bibfnamefont{Y.-D.} \bibnamefont{Li}},
  \bibinfo{journal}{Chin. Phys. Lett.} \textbf{\bibinfo{volume}{30}},
  \bibinfo{pages}{011201} (\bibinfo{year}{2013}).

\bibitem[{\citenamefont{Azevedo et~al.}(2020)\citenamefont{Azevedo,
  Gon\c{c}alves, and Moreira}}]{Azevedo:2019hqp}
\bibinfo{author}{\bibfnamefont{C.}~\bibnamefont{Azevedo}},
  \bibinfo{author}{\bibfnamefont{V.~P.} \bibnamefont{Gon\c{c}alves}},
  \bibnamefont{and} \bibinfo{author}{\bibfnamefont{B.~D.}
  \bibnamefont{Moreira}}, \bibinfo{journal}{Phys. Rev. C}
  \textbf{\bibinfo{volume}{101}}, \bibinfo{pages}{024914}
  (\bibinfo{year}{2020}), \eprint{1911.10861}.

\bibitem[{\citenamefont{Yu et~al.}(2022)\citenamefont{Yu, Zhao, Cai, Gao, Hu,
  and Yang}}]{Yu:2022hdt}
\bibinfo{author}{\bibfnamefont{G.}~\bibnamefont{Yu}},
  \bibinfo{author}{\bibfnamefont{Z.}~\bibnamefont{Zhao}},
  \bibinfo{author}{\bibfnamefont{Y.}~\bibnamefont{Cai}},
  \bibinfo{author}{\bibfnamefont{Q.}~\bibnamefont{Gao}},
  \bibinfo{author}{\bibfnamefont{Q.}~\bibnamefont{Hu}}, \bibnamefont{and}
  \bibinfo{author}{\bibfnamefont{H.}~\bibnamefont{Yang}}
  (\bibinfo{year}{2022}), \eprint{2209.11439}.

\bibitem[{\citenamefont{V.W.~Hughes}(1971)}]{Hughes:1971}
\bibinfo{author}{\bibfnamefont{B.~M.} \bibnamefont{V.W.~Hughes}},
  \bibinfo{journal}{Bull.Am.Phys.Soc.} \textbf{\bibinfo{volume}{16}},
  \bibinfo{pages}{65} (\bibinfo{year}{1971}).

\bibitem[{\citenamefont{Nemenov}(1972)}]{Nemenov:1972}
\bibinfo{author}{\bibfnamefont{L.}~\bibnamefont{Nemenov}},
  \bibinfo{journal}{Sov.J.Nucl.Phys.} \textbf{\bibinfo{volume}{15}},
  \bibinfo{pages}{582} (\bibinfo{year}{1972}).

\bibitem[{\citenamefont{Cid~Vidal et~al.}(2019)\citenamefont{Cid~Vidal, Ilten,
  Plews, Shuve, and Soreq}}]{CidVidal:2019qub}
\bibinfo{author}{\bibfnamefont{X.}~\bibnamefont{Cid~Vidal}},
  \bibinfo{author}{\bibfnamefont{P.}~\bibnamefont{Ilten}},
  \bibinfo{author}{\bibfnamefont{J.}~\bibnamefont{Plews}},
  \bibinfo{author}{\bibfnamefont{B.}~\bibnamefont{Shuve}}, \bibnamefont{and}
  \bibinfo{author}{\bibfnamefont{Y.}~\bibnamefont{Soreq}},
  \bibinfo{journal}{Phys. Rev. D} \textbf{\bibinfo{volume}{100}},
  \bibinfo{pages}{053003} (\bibinfo{year}{2019}), \eprint{1904.08458}.

\bibitem[{\citenamefont{Moffat}(1975)}]{Moffat:1975uw}
\bibinfo{author}{\bibfnamefont{J.~W.} \bibnamefont{Moffat}},
  \bibinfo{journal}{Phys. Rev. Lett.} \textbf{\bibinfo{volume}{35}},
  \bibinfo{pages}{1605} (\bibinfo{year}{1975}).

\bibitem[{\citenamefont{Brodsky and Lebed}(2009)}]{Brodsky:2009gx}
\bibinfo{author}{\bibfnamefont{S.~J.} \bibnamefont{Brodsky}} \bibnamefont{and}
  \bibinfo{author}{\bibfnamefont{R.~F.} \bibnamefont{Lebed}},
  \bibinfo{journal}{Phys. Rev. Lett.} \textbf{\bibinfo{volume}{102}},
  \bibinfo{pages}{213401} (\bibinfo{year}{2009}), \eprint{0904.2225}.

\bibitem[{\citenamefont{Gargiulo
  et~al.}(2024{\natexlab{a}})\citenamefont{Gargiulo, Palmisano, Di~Meco,
  Diociaiuti, Sarra, and Paesani}}]{Gargiulo:2023tci}
\bibinfo{author}{\bibfnamefont{R.}~\bibnamefont{Gargiulo}},
  \bibinfo{author}{\bibfnamefont{S.}~\bibnamefont{Palmisano}},
  \bibinfo{author}{\bibfnamefont{E.}~\bibnamefont{Di~Meco}},
  \bibinfo{author}{\bibfnamefont{E.}~\bibnamefont{Diociaiuti}},
  \bibinfo{author}{\bibfnamefont{I.}~\bibnamefont{Sarra}}, \bibnamefont{and}
  \bibinfo{author}{\bibfnamefont{D.}~\bibnamefont{Paesani}},
  \bibinfo{journal}{J. Phys. G} \textbf{\bibinfo{volume}{51}},
  \bibinfo{pages}{045004} (\bibinfo{year}{2024}{\natexlab{a}}),
  \eprint{2309.11683}.

\bibitem[{\citenamefont{Gargiulo
  et~al.}(2024{\natexlab{b}})\citenamefont{Gargiulo, Di~Meco, and
  Palmisano}}]{Gargiulo:2024zyc}
\bibinfo{author}{\bibfnamefont{R.}~\bibnamefont{Gargiulo}},
  \bibinfo{author}{\bibfnamefont{E.}~\bibnamefont{Di~Meco}}, \bibnamefont{and}
  \bibinfo{author}{\bibfnamefont{S.}~\bibnamefont{Palmisano}}
  (\bibinfo{year}{2024}{\natexlab{b}}), \eprint{2409.11342}.

\bibitem[{\citenamefont{Ji and Lamm}(2018)}]{Ji:2017lyh}
\bibinfo{author}{\bibfnamefont{Y.}~\bibnamefont{Ji}} \bibnamefont{and}
  \bibinfo{author}{\bibfnamefont{H.}~\bibnamefont{Lamm}},
  \bibinfo{journal}{Phys. Rev. D} \textbf{\bibinfo{volume}{98}},
  \bibinfo{pages}{053008} (\bibinfo{year}{2018}), \eprint{1706.04986}.

\bibitem[{\citenamefont{Fael and Mannel}(2018)}]{Fael:2018ktm}
\bibinfo{author}{\bibfnamefont{M.}~\bibnamefont{Fael}} \bibnamefont{and}
  \bibinfo{author}{\bibfnamefont{T.}~\bibnamefont{Mannel}},
  \bibinfo{journal}{Nucl. Phys. B} \textbf{\bibinfo{volume}{932}},
  \bibinfo{pages}{370} (\bibinfo{year}{2018}), \eprint{1803.08880}.

\bibitem[{\citenamefont{Fadin et~al.}(1990)\citenamefont{Fadin, Khoze, and
  Sjostrand}}]{Fadin:1990wx}
\bibinfo{author}{\bibfnamefont{V.~S.} \bibnamefont{Fadin}},
  \bibinfo{author}{\bibfnamefont{V.~A.} \bibnamefont{Khoze}}, \bibnamefont{and}
  \bibinfo{author}{\bibfnamefont{T.}~\bibnamefont{Sjostrand}},
  \bibinfo{journal}{Z. Phys. C} \textbf{\bibinfo{volume}{48}},
  \bibinfo{pages}{613} (\bibinfo{year}{1990}).

\bibitem[{\citenamefont{Bogomyagkov et~al.}(2018)\citenamefont{Bogomyagkov,
  Druzhinin, Levichev, Milstein, and Sinyatkin}}]{Bogomyagkov:2017uul}
\bibinfo{author}{\bibfnamefont{A.}~\bibnamefont{Bogomyagkov}},
  \bibinfo{author}{\bibfnamefont{V.}~\bibnamefont{Druzhinin}},
  \bibinfo{author}{\bibfnamefont{E.}~\bibnamefont{Levichev}},
  \bibinfo{author}{\bibfnamefont{A.}~\bibnamefont{Milstein}}, \bibnamefont{and}
  \bibinfo{author}{\bibfnamefont{S.}~\bibnamefont{Sinyatkin}},
  \bibinfo{journal}{EPJ Web Conf.} \textbf{\bibinfo{volume}{181}},
  \bibinfo{pages}{01032} (\bibinfo{year}{2018}), \eprint{1708.05819}.

\bibitem[{\citenamefont{Ablikim et~al.}(2013{\natexlab{a}})}]{BESIII:2013csc}
\bibinfo{author}{\bibfnamefont{M.}~\bibnamefont{Ablikim}} \bibnamefont{et~al.}
  (\bibinfo{collaboration}{BESIII}), \bibinfo{journal}{Phys. Rev. D}
  \textbf{\bibinfo{volume}{88}}, \bibinfo{pages}{032007}
  (\bibinfo{year}{2013}{\natexlab{a}}), \eprint{1307.1189}.

\bibitem[{\citenamefont{Ablikim et~al.}(2024{\natexlab{a}})}]{BESIII:2021ocn}
\bibinfo{author}{\bibfnamefont{M.}~\bibnamefont{Ablikim}} \bibnamefont{et~al.}
  (\bibinfo{collaboration}{BESIII}), \bibinfo{journal}{Phys. Rev. D}
  \textbf{\bibinfo{volume}{109}}, \bibinfo{pages}{052006}
  (\bibinfo{year}{2024}{\natexlab{a}}), \eprint{2111.13881}.

\bibitem[{\citenamefont{Ablikim et~al.}(2013{\natexlab{b}})}]{BESIII:2012lxx}
\bibinfo{author}{\bibfnamefont{M.}~\bibnamefont{Ablikim}} \bibnamefont{et~al.}
  (\bibinfo{collaboration}{BESIII}), \bibinfo{journal}{Phys. Rev. D}
  \textbf{\bibinfo{volume}{87}}, \bibinfo{pages}{032003}
  (\bibinfo{year}{2013}{\natexlab{b}}), \eprint{1208.1461}.

\bibitem[{\citenamefont{Petrelli et~al.}(1998)\citenamefont{Petrelli, Cacciari,
  Greco, Maltoni, and Mangano}}]{Petrelli:1997ge}
\bibinfo{author}{\bibfnamefont{A.}~\bibnamefont{Petrelli}},
  \bibinfo{author}{\bibfnamefont{M.}~\bibnamefont{Cacciari}},
  \bibinfo{author}{\bibfnamefont{M.}~\bibnamefont{Greco}},
  \bibinfo{author}{\bibfnamefont{F.}~\bibnamefont{Maltoni}}, \bibnamefont{and}
  \bibinfo{author}{\bibfnamefont{M.~L.} \bibnamefont{Mangano}},
  \bibinfo{journal}{Nucl. Phys. B} \textbf{\bibinfo{volume}{514}},
  \bibinfo{pages}{245} (\bibinfo{year}{1998}), \eprint{hep-ph/9707223}.

\bibitem[{\citenamefont{Sommerfeld}(1931)}]{Sommerfeld:1931qaf}
\bibinfo{author}{\bibfnamefont{A.}~\bibnamefont{Sommerfeld}},
  \bibinfo{journal}{Annalen Phys.} \textbf{\bibinfo{volume}{403}},
  \bibinfo{pages}{257} (\bibinfo{year}{1931}).

\bibitem[{\citenamefont{Sakharov}(1948)}]{Sakharov:1948plh}
\bibinfo{author}{\bibfnamefont{A.~D.} \bibnamefont{Sakharov}},
  \bibinfo{journal}{Zh. Eksp. Teor. Fiz.} \textbf{\bibinfo{volume}{18}},
  \bibinfo{pages}{631} (\bibinfo{year}{1948}).

\bibitem[{\citenamefont{Schwinger}(1973)}]{Schwinger:1973rv}
\bibinfo{author}{\bibfnamefont{J.~S.} \bibnamefont{Schwinger}},
  \emph{\bibinfo{title}{{PARTICLES, SOURCES AND FIELDS. VOLUME II}}}
  (\bibinfo{year}{1973}).

\bibitem[{\citenamefont{Fadin and Khoze}(1987)}]{Fadin:1987wz}
\bibinfo{author}{\bibfnamefont{V.~S.} \bibnamefont{Fadin}} \bibnamefont{and}
  \bibinfo{author}{\bibfnamefont{V.~A.} \bibnamefont{Khoze}},
  \bibinfo{journal}{JETP Lett.} \textbf{\bibinfo{volume}{46}},
  \bibinfo{pages}{525} (\bibinfo{year}{1987}).

\bibitem[{\citenamefont{Navas et~al.}(2024)}]{ParticleDataGroup:2024cfk}
\bibinfo{author}{\bibfnamefont{S.}~\bibnamefont{Navas}} \bibnamefont{et~al.}
  (\bibinfo{collaboration}{Particle Data Group}), \bibinfo{journal}{Phys. Rev.
  D} \textbf{\bibinfo{volume}{110}}, \bibinfo{pages}{030001}
  (\bibinfo{year}{2024}).

\bibitem[{\citenamefont{Ablikim et~al.}(2022)}]{BESIII:2021cxx}
\bibinfo{author}{\bibfnamefont{M.}~\bibnamefont{Ablikim}} \bibnamefont{et~al.}
  (\bibinfo{collaboration}{BESIII}), \bibinfo{journal}{Chin. Phys. C}
  \textbf{\bibinfo{volume}{46}}, \bibinfo{pages}{074001}
  (\bibinfo{year}{2022}), \eprint{2111.07571}.

\bibitem[{\citenamefont{Ablikim et~al.}(2024{\natexlab{b}})}]{BESIII:2024lks}
\bibinfo{author}{\bibfnamefont{M.}~\bibnamefont{Ablikim}} \bibnamefont{et~al.}
  (\bibinfo{collaboration}{BESIII}), \bibinfo{journal}{Chin. Phys. C}
  \textbf{\bibinfo{volume}{48}}, \bibinfo{pages}{093001}
  (\bibinfo{year}{2024}{\natexlab{b}}), \eprint{2403.06766}.

\bibitem[{\citenamefont{Achasov et~al.}(2024)}]{Achasov:2023gey}
\bibinfo{author}{\bibfnamefont{M.}~\bibnamefont{Achasov}} \bibnamefont{et~al.},
  \bibinfo{journal}{Front. Phys. (Beijing)} \textbf{\bibinfo{volume}{19}},
  \bibinfo{pages}{14701} (\bibinfo{year}{2024}), \eprint{2303.15790}.

\bibitem[{\citenamefont{Ablikim et~al.}(2010)}]{BESIII:2009fln}
\bibinfo{author}{\bibfnamefont{M.}~\bibnamefont{Ablikim}} \bibnamefont{et~al.}
  (\bibinfo{collaboration}{BESIII}), \bibinfo{journal}{Nucl. Instrum. Meth. A}
  \textbf{\bibinfo{volume}{614}}, \bibinfo{pages}{345} (\bibinfo{year}{2010}),
  \eprint{0911.4960}.

\end{thebibliography}

\end{document}